\documentclass[aps,pra,reprint,superscriptaddress]{revtex4-2}
\usepackage{graphicx}
\usepackage{color}
\usepackage{amsmath,amssymb}
\usepackage{bm}
\usepackage{upgreek}
\usepackage{enumitem}
\usepackage[colorlinks,urlcolor=blue,citecolor=blue,linkcolor=blue]{hyperref}

\usepackage{bbold}

\usepackage[english]{babel}

\begin{document}

\title{Photoelectric detection of single spins in diamond by optically controlled discharge of long-lived trap states}

\author{A. C. Ulibarri}
\affiliation{School of Physics, University of Melbourne, Victoria 3010, Australia}
\author{D. J. McCloskey}
\affiliation{School of Physics, University of Melbourne, Victoria 3010, Australia}
\affiliation{Australian Research Council Centre of Excellence in Quantum Biophysics, School of Physics, University of Melbourne, Victoria 3010, Australia}
\author{D. Wang}
\affiliation{School of Physics, University of Melbourne, Victoria 3010, Australia}
\author{N. Dontschuk}
\affiliation{School of Physics, University of Melbourne, Victoria 3010, Australia}
\author{A. M. Martin}
\affiliation{School of Physics, University of Melbourne, Victoria 3010, Australia}
\author{A.~A.~Wood}
\email{alexander.wood@unimelb.edu.au}
\affiliation{School of Physics, University of Melbourne, Victoria 3010, Australia}

\date{\today}
\begin{abstract}\label{abstract}

\noindent Electrical detection methods for solid-state spins are attractive for quantum technologies, being readily chip-scalable and not subject to the small photon budgets of single emitters. However, realising electrical spin readout in wide-bandgap materials with similar fidelity and bandwidth to optical approaches remains challenging. Here, we introduce a photoelectrical spin readout scheme that detects spin information stored long-term as trapped electrical charges. Using nitrogen-vacancy (NV) centres in diamond as a model system, spin-dependent photoionisation generates charge carriers that are stored in long-lived trap states at a diamond-metal Schottky junction. On-demand illumination of the junction under electrical bias releases stored charge, yielding a photocurrent transient proportional to the amount of trapped charge and hence spin state. Spin readout after coherent control of single NVs is demonstrated using charge readout in a protocol we call charge-capture detected magnetic resonance (CCDMR), and we use charge-based imaging to identify charge carrier generation and trapping processes. Our results establish CCDMR as a new technique for solid-state spin qubit readout, combining attaractive features of electrical detection with the stability of long-lived charge traps in wide-bandgap materials.

\end{abstract}
\maketitle

\section{Introduction}\label{sec: intro}

\noindent Optical preparation and readout methods confer myriad advantages to spin qubits based on colour centres in wide-bandgap (WBG) semiconductors, such as fast, spectrally-selective initialisation and detection, room-temperature operation, and simple experimental setups based on commercial electro-optical components. However, optical readout of spin qubits comes with the drawback of typically low photon collection efficiency~\cite{barry_sensitivity_2020} which is further exacerbated by the high refractive index of WBG materials such as diamond and silicon carbide~\cite{yeung_anti-reflection_2012}. Further, optical detection schemes require bulky, high-numerical aperture collection optics that pose challenges to integration and miniaturisation of solid state spin qubits into practical sensing or information processing devices. While integrated photonic structures such as solid immersion lenses~\cite{jamali_microscopic_2014, christinck_bright_2023} or waveguides~\cite{sipahigil_integrated_2016, bradac_quantum_2019} can significantly increase photon collection efficiencies~\cite{katsumi_recent_2025}, alternative readout mechanisms such as photoelectric detection (PD)~\cite{bourgeois_photoelectric_2015, bourgeois_photoelectric_2020} combine the simplicity and flexibility of optical probing with the near-unity collection efficiency of electrical charge carriers produced by defects inside the host material. PD has been demonstrated for single centres and defect ensembles in both diamond~\cite{bourgeois_photoelectric_2015,siyushev_photoelectrical_2019} and silicon carbide~\cite{niethammer_coherent_2019, nishikawa_coherent_2025} and has been employed to detect nuclear spins~\cite{gulka_room-temperature_2021,morishita_room_2020} and, more recently, proximal electron spins~\cite{rubinas_electrical_2025}. \newline

\noindent Photoelectrical readout schemes have experimental challenges, such as the presence of background photocurrent stemming from other defects~\cite{hruby_magnetic_2022} or from the metal-semiconductor junctions required for electric interfacing~\cite{rieger_fast_2024}, limited measurement bandwidth owing to slow current amplifiers~\cite{hrubesch_efficient_2017, gulka_pulsed_2017}, trapping of charge carriers by other defects~\cite{jayakumar_optical_2016, lozovoi_imaging_2022, wood_3d-mapping_2024}, space-charge screening~\cite{lozovoi_probing_2020, goldblatt_quantum_2024}, competing ionisation and recombination processes~\cite{bourgeois_enhanced_2017, bourgeois_photoelectric_2022} and crosstalk between metal electrodes and the electromagnetic fields used for spin qubit control~\cite{wirtitsch_microelectronic_2024}. Many of the issues encountered in photoelectric detection are directly related to unanswered questions pertaining to the impacts of charge carrier generation, transport, and trapping on the electronic properties of WBG materials across different length scales~\cite{lozovoi_optical_2021, lozovoi_detection_2023,delord_correlated_2024, ji_correlated_2024,monge_beyond_2025}, and are also relevant to the stabilisation of colour centre charge states~\cite{gardill_probing_2021, gorlitz_coherence_2022, wood_room-temperature_2023, zhang_neutral_2023, garcia-arellano_photo-induced_2024}. 

\begin{figure*}[t]
\centering
  \includegraphics[width=\textwidth]{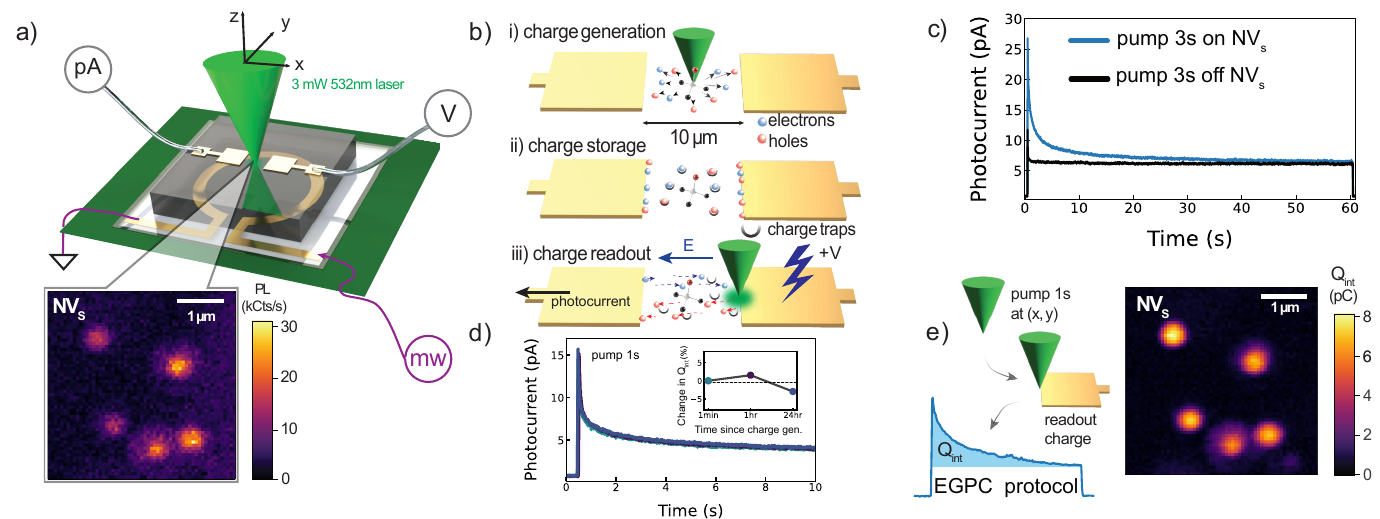}
  \caption{{\bf Experimental schematic and charge readout procedure.} a) A diamond sample with metal surface contacts is connected to a voltage source and a current preamplifier for photocurrent measurements. The sample is positioned in a scanning confocal microscope for optical excitation. Inset: A PL map of a region containing 5~NV centres which were employed in this work. b) Trapped charge detection mechanism. i) 532\,nm laser excitation generates a diffusing distribution of electrons and holes from a single NV centre. ii) Photo-generated charges become trapped in long-lived states residing within the diamond and at diamond-metal interfaces. iii) A bias voltage is applied between the electrodes while a laser excites the diamond-metal interface, generating a photocurrent transient proportional to the total number of trapped photocarriers. c) Photocurrent transients for laser pumping a single NV centre for 3\,s at 3.5\,mW (blue) compared to the same photo-pumping parameters at an adjacent empty spot (black). d) Photocurrent transients for wait times after pumping of 1 minute, 1 hour and 24 hours (turquoise, purple, and blue respectively). Inset: change in total integrated charge measured for increasing delay periods. e) The excess photocurrent is integrated to obtain a measure of the total trapped charge produced by the NV centre during laser pumping. Performing this measurement repeatedly as the pump laser spot is scanned across the sample enables photoelectric imaging of single NV centres.}
  \label{fig:1}
\end{figure*}

\noindent In this work, we introduce a new scheme for PD of single spins in WBG materials based on carrier generation from spin defects and storage of the resulting charge in highly stable interface trap states. Charge readout is then performed by remote photoelectrical readout with laser illumination of the semiconductor-metal interface, rather than the spin defect itself or any trapped bulk charge distributions. We use single nitrogen-vacancy (NV) centre defects in a high-purity CVD diamond as a prototypical example and demonstrate imaging of single defects based on liberation of trapped photocarriers. The photocurrent generated by illumination of the metal-diamond interface exhibits a characteristic amplitude that reveals the magnitude of NV-generated charge, and spin-dependent carrier generation from the NV then allows us to obtain an electron spin resonance spectrum from trapped charge readout. We study this process in detail and support our identification of electrode-generated hole photocurrent as the underlying readout and charge-liberating mechanism by tracking carrier transport via hole capture by single defect centres. Our results make important progress in understanding the charge environment inside wide-bandgap materials interfaced with electrical readout infrastructure and present an alternative method to conventional photoelectric detection with several attractive additional features. While we use the archetypal NV centre in diamond for our present work, our method can be easily extended to other defects and host materials.

\section{Experimental Section}\label{sec: experiment}

\noindent Our experiment and the essential elements of our new PD method, depicted in Fig.~\ref{fig:1}(a), consists of a custom scanning confocal microscopy system simultaneously capable of standard photoluminescence (PL) measurements and PD. Planar $10\,\upmu\text{m}^2$ Cr/Au electrodes are fabricated onto the oxygen-terminated surface of an electron-irradiated $\langle100\rangle$-cut electronic grade CVD diamond (Element 6) by photolithography. A 0.8~NA apochromatic air objective focuses light to a diffraction limited ($\sim$400\,nm) spot and collects NV$^-$ photoluminescence (650-750\,nm) with a single-photon counting module (Excelitas). Within the 10$\,\upmu$m lateral inter-electrode gap are individually resolvable NV centres with an average separation distance of 1-2\,$\upmu$m, distributed uniformly throughout the bulk of the sample. An example photoluminescence map is shown in Fig.~\ref{fig:1}(a) for five NV centres situated approximately 4\,$\upmu$m below the diamond surface. The continuous-wave illumination is tunable from 500-780\,nm, though we typically employ 532\,nm green light for optical polarisation of NV centres together with 633\,nm for charge state control and 594\,nm for NV charge state readout~\cite{aslam_photo-induced_2013}. Bias voltages (-10 to 10\,V) are applied to the metal electrodes, and one electrode is connected to a high gain current amplifier (Stanford Research Systems SR570, nominal gain of 20\,pA/V), the output of which is low-pass filtered (37\,Hz cutoff) to eliminate 50\,Hz line noise pickup on the electrodes.

\section{Results}

\subsection{Charge-based readout protocol}

\noindent An archetypal experiment proceeds as follows: we begin by illuminating a single NV centre under zero applied bias voltage for some time with green light (532\,nm) at powers sufficient to drive cycling between the negative (NV$^-$) and neutral (NV$^0$) charge states of the defect~\cite{aslam_photo-induced_2013}. With each successive cycle of ionisation (NV$^{-}$$\rightarrow$NV$^0$) and recombination (NV$^0$$\rightarrow$NV$^-$), free electrons and holes are injected into the conduction and valence band of the diamond respectively. While these free carriers can be quickly trapped by a host of intrinsic impurities, such as donor-like substitutional nitrogen (N)~\cite{jayakumar_optical_2016,ashfold_nitrogen_2020, bian_nanoscale_2021, zheng_coherence_2022, wood_3d-mapping_2024}, acceptor-like boron (B)~\cite{grunwald_photoconductive_2021, schreck_charge_2020}, and a variety of surface states~\cite{itoh_trapping_2006, stacey_evidence_2019, barson_nanoscale_2021, trofimov_voltage_2025}, they can also be trapped in states localised to the diamond-metal electrode interface~\cite{nicholls_role_2019, lew_investigation_2020}. At low N and B doping densities ($<$10\,ppm), diamond, like other WBG materials hosting quantum defects at room temperature or below, exhibits negligible free carriers due to thermal excitation, resulting in trapped charges being stored essentially indefinitely~\cite{dhomkar_long-term_2016, monge_reversible_2023}.\newline

\noindent The stored charge is then read out via an unconventional mechanism. The metal-diamond-metal interface is electrically equivalent to two back-to-back Schottky diodes~\cite{grillo_currentvoltage_2021} which prevents direct charge injection from the electrodes, but current can be generated through optical illumination. We apply a bias voltage between the electrodes, then shine laser light at the forward-biased diamond-electrode interface (right-hand electrode) as illustrated in Fig.~\ref{fig:1}(b). This elicits an electrode-generated photocurrent (EGPC), an example of which is shown in Fig.\ref{fig:1}(c) for an applied bias of 2.2\,V and 3.5\,mW of 532\,nm laser illumination. The EGPC originates from the same mechanism responsible for conventional metal-semiconductor photoconductivity: light penetrates into the metal and excites electronic transitions within the film, generating free holes~\cite{sze_physics_2006, kukushkin_visible_2019}. These holes quickly diffuse, and when the metal thickness approaches the hole diffusion length (10\,nm for Cr/Au~\cite{kukushkin_visible_2019}) -- at the interface with the diamond -- they can be injected into the diamond where they are swept towards the opposite electrode by the applied electric field. 

\subsection{Mapping NV charge generation}

\noindent While illumination of the electrode junction always creates an EGPC, pumping the NV beforehand yields a sharp, characteristic photocurrent transient signal, initially peaking before decaying over several seconds to a baseline value. The very short time transient behaviour ($<\,$10\,ms) is obscured by the limited detection bandwidth of the current amplifier and filter, but a slow decay transient is unaffected. The additional signal originates from past charge cycling of the source NV centre, as demonstrated in Fig.~\ref{fig:1}(c) where illumination adjacent to, but not on, an NV yields no transient (black trace). We interpret the increased signal as arising from laser-induced discharging of trapped charges generated during the NV centre pumping step. Furthermore, for a single illumination time/power combination applied to a given NV centre, the transient decay remains the same over extremely long timescales of at least a day (Fig.\ref{fig:1}(d)). We calculate the integrated additional charge $Q_{\text{int}}$ released during the readout step by subtracting the steady-state EGPC from the transient signal, as shown in Fig. \ref{fig:1}(e). To demonstrate definitively the role played by NV charge pumping, we illuminate with the green laser ($P = 3.5\,$mW, $t = 1$\,s) a point $(x,y)$ and then readout the accumulated charge by applying a bias voltage and illuminating at the metal-diamond junction. Scanning across the same spatial region as in Fig. \ref{fig:1}(a) results in a $Q_{\text{int}}$ vs laser position map in Fig.~\ref{fig:1}(e), with single NV centres identified as the domainat sources of additional charge. Notably, photoelectrical imaging demonstrates a 30\% improvement in spatial resolution and 10\% contrast improvement compared to the equivalent fluorescence image.

\subsection{Charge-capture detected magnetic resonance}\label{subsec: CCDMR}

\begin{figure}[t]
    \centering
    \includegraphics[width=\columnwidth]{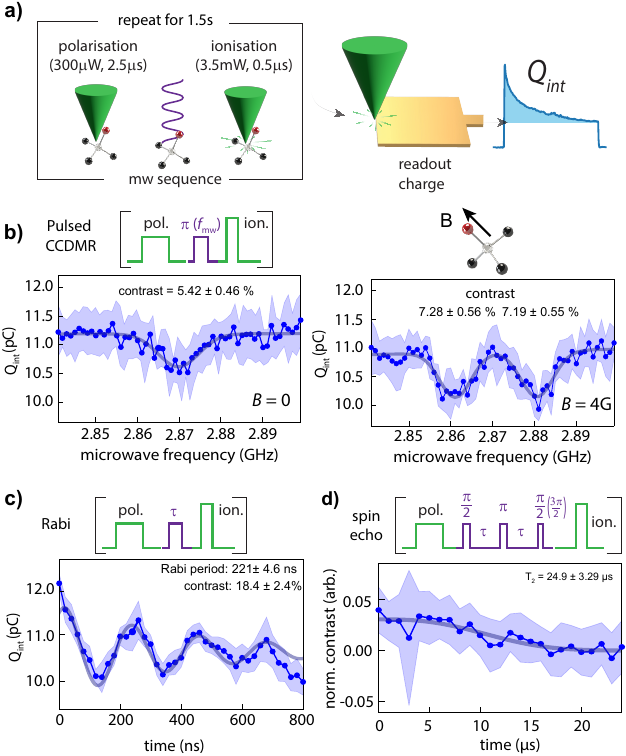}
    \caption{{\bf Spin measurement and coherent control with CCDMR.} a) Schematic showing measurement process, a laser polarisation pulse (300\,$\upmu$W for 2.5\,$\upmu$s) initiates an $m = 0$ spin state before a microwave pulse is applied, and the resulting spin state is converted to charges via an ionisation pulse (3.5\,mW for 500\,ns). The whole sequence is repeated for 1.5~s before readout of the charges is performed at the electrode. b) (left): CCDMR showing characteristic NV zero-field resonance at 2.87\,GHz and (right) CCDMR under a bias magnetic field of 4\,G oriented along the NV axis. c) Rabi oscillation measured at 2.87\,GHz under zero bias field. d) Hahn-echo signal measured for the lower energy resonance ($f = 2860\,$MHz) shown in panel b). Shaded regions are one standard deviation in measurement outcomes across 10 repeats.}
    \label{fig:2}
\end{figure}

\noindent The photoionisation probability of negatively-charged NV centres is spin-dependent due to the increased probability of the magnetic states $m_S = \pm1$ to cross to a coupled metastable singlet state when compared to the $m_S = 0$ state, which reduces the ionisation rate from the excited $^3E$ state~\cite{doherty_nitrogen-vacancy_2013, hopper_near-infrared-assisted_2016,bourgeois_photoelectric_2020}. Analogous to photoelectrically-detected magnetic resonance~\cite{bourgeois_photoelectric_2015}, spin-to charge conversion~\cite{shields_efficient_2015} and ancilla-assisted charge measurements~\cite{jayakumar_long-term_2020}, it follows that NV spin information should be encoded into the EGPC transient signal. We adapted a standard optically-detected magnetic resonance pulse sequence for use with EGPC readout, a protocol we term \emph{charge-capture detected magnetic resonance}, or CCDMR. Briefly, a short (2.5\,$\upmu$s) 532\,nm pulse at 300\,$\upmu$W is used to initialise a single NV$^-$ defect into the $m_S = 0$ ground state. Low powers are chosen to minimise unwanted carrier generation. Then, a $100\,$ns microwave $\pi$-pulse of varying frequency is applied and a second, shorter (500\,ns) 532\,nm pulse of high power (3.5\,mW) is then applied to the NV to initiate photoionisation. The sequence is repeated hundreds of thousands of times (over approximately 1.5\,s) to generate a spin-dependent trapped charge population in the sample. Finally, we read the trapped charge by illuminating the positively-biased electrode edge with 3.5\,mW of green at a bias voltage of 2.2\,V as discussed previously to obtain a measure of the accumulated charge for each microwave frequency. We observe a dip in the CCDMR spectrum at 2.87\,GHz which exhibits a contrast of 5.42$\,\pm{\,0.46}$\,$\%$ in Fig.~\ref{fig:2}(b,left). The negative dip indicates a reduction in NV-generated photocarriers at resonance, consistent with a reduced rate of NV$^{-/0}$ charge cycling due to shelving by the longer-lived spin singlet manifold. Applying a bias magnetic field of 4\,G along the NV axis splits the CCDMR resonance, shown in Fig.~\ref{fig:2}(b, right), by 20\,MHz and results in a contrast of 7.28$\,\pm{\,0.56}$\,$\%$ and 7.19$\,\pm{\,0.55}$\,$\%$, facilitating independent addressing of each spin projection. \newline

\noindent We then verified our readout mechanism is compatible with coherent spin control protocols by demonstrating Rabi oscillations with CCDMR. Using the same initialisation, ionisation, and EGPC readout pulses as described previously, we vary the duration of the microwave pulse while resonant with the zero field CCMDR peak shown in Fig.~\ref{fig:2}(b) and observe unambiguous Rabi oscillations with a period of 221.1\,$\pm{\,4.6}$ns. Finally, we performed a spin-echo measurement using the same charge readout protocol as before. We measured the total integrated charge for spin-echo sequences terminated with either a $\pi$/2 or 3$\pi$/2 projection pulse to measure a normalised spin-echo profile and extract a $T_2$ decay time of 24.90\,$\pm{\,3.29}$\,$\upmu$s. We compared all spin measurements to standard optical techniques and obtained similar values for the Rabi period and $T_2$.\newline
\begin{figure}[t]
    \centering
    \includegraphics[width=\columnwidth]{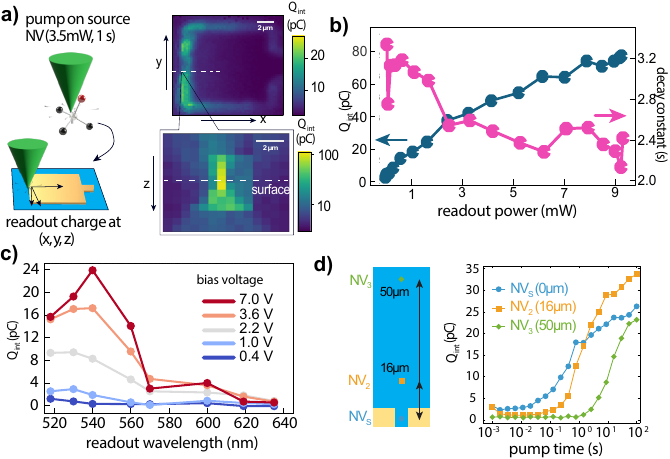}
    \caption{\textbf{Optical power, wavelength and spatial dependence of charge capture and readout scheme.} a) Integrated charge as a function of space over the positively-biased electrode pad at 2.2~V. The inset shows an x-z slice along the contact left edge. b) Left axis demonstrates integrated charge as a function of green illumination power at the contact to induce EGPC. Right axis depicts the EGPC limiting time constant. c) Integrated charge measured as a function of applied bias and contact illumination wavelength for a 532~nm NV pump of 1~s at 3.5~mW. d) Schematic of three target NV centres and their corresponding integrated charge as a function of pump time on the NV.}
    \label{fig:3}
\end{figure}

\noindent Taken together, these proof-of-principle results show that CCDMR constitutes a photoelectric spin-state readout method with fidelity comparable to standard optical and PD techniques. The obtained Rabi contrast of 18.4\,$\pm{\,2.4}$\,$\%$ even exceeds that typically obtained by PDMR measurements ~\cite{bourgeois_photoelectric_2020}, which may suggest more efficient suppression of background charge sources in this delayed-readout configuration. However, the significant noise in CCDMR measurements needs to be better understood, and we also note that our current CCDMR measurement sequences are significantly slower than optical equivalents owing to the requirement for $\sim$\,10\,s of illumination to fully liberate all NV-generated trapped charges. As these issues are the primary bottleneck in limiting applicability of our technique in NV-based sensing and information processing, we now turn our attention to more detailed examination of the readout mechanism at play.

\subsection{Readout mechanism}\label{subsec: interface}
\noindent Optical illumination of the electrode-diamond junction serves to liberate stored charge originating from single NV centres, and we have shown that spin information can be extracted from the accumulated charge. We now focus on the readout mechanism in more detail. First, we show that illumination of the metal-diamond interface is the origin of both the EGPC and the liberation of trapped NV centre charge by probing the spatial dependence of the readout process. To this end, we pump a source NV centre (NV$_\text{S}$ in Fig.~\ref{fig:1}(a)) with 3.5\,mW of green light for 1\,s to generate charges, then vary the position where the readout laser pulse is applied across the positively-biased contact. We also change the focus position into the diamond to examine the depth dependence of the readout process.\newline

\noindent Electrode photocurrent was found to be strongest at the contact edge facing the ground electrode in Fig.~\ref{fig:3}(a), and exhibited a strong dependence in depth, showing that the highest integrated charge was measured at the diamond-electrode boundary $\approx$\,4.5\,$\upmu$m above the source NV. Next, we fixed the position of the readout laser and depth to yield maximum signal and varied the power of the 532\,nm illumination at the contact to probe the dependence of the integrated charge and temporal dynamics of the EGPC transient. For a single NV pump power and duration (3.5\,mW, and 1\,s respectively) increasing power at the electrode exhibited increasing integrated charge (as seen in Fig.~\ref{fig:3}(b) left axis) with a tendency towards saturation level at high power. Further, the EGPC transient was fit with a double exponential and the limiting (slow) decay constant was found to monotonically decrease as the laser power increased (as seen in Fig.~\ref{fig:3}(c) right axis). This follows intuitively, as higher power illumination should liberate trapped charge faster. Illuminating the contact with higher powers appears to lead to larger integrated charge before eventually saturating, indicating that the more power used, the more charge is released. Ideally the laser power would be set to the power saturation point to completely liberate the NV generated trapped charge. However, powers in the 10\,s of mW range are difficult to implement and could also result in damage to the electrodes with tight focusing.\newline

\noindent Finally, the wavelength of illumination light used for readout to probe the energetics of the associated trapping defects. We found that for the same source NV pumping parameters (1\,s and 3.5\,mW at  532\,nm) and independent of applied bias voltage, the measured integrated charge increased significantly for wavelengths below 570\,nm (photon energies of 2.2\,eV) as shown in Fig.~\ref{fig:3}(d), with a maximum at 540\,nm (2.3\,eV). Curiously, 2.2\,eV is also the observed photoionisation threshold of substitutional nitrogen~\cite{bourgeois_enhanced_2017}, though the spatial dependence of the readout illumination strongly suggests that charge storage occurs at interface states introduced during the Au/Cr metallisation of the diamond surface.\newline

\noindent When the bias potential is inverted, the spatial dependence observed above is replicated on the left-hand electrode, and no EGPC or charge transient is observed when the readout laser targets the right-hand (negatively biased) electrode, consistent with our model of EGPC discussed earlier. Along the contact facing the ground pad, there is no significant spatial dependence of $Q_{\text{int}}$, suggesting either that the interface states are uniformly distributed or that charge transfer is limited by a global potential barrier rather than local variations. In other experiments, we found that illumination of the electrode edge without an applied bias voltage following the generation of charges from a source NV attenuates the integrated charge detected in subsequent readout, indicating that light alone is sufficient to deplete trapped charges.\newline

\noindent We then studied the temporal accumulation of trapped charges as a function of the source NV distance to the electrode. We located three NV centres, NV$_S$ (the source NV from previous experiments), located almost exactly in the centre of the electrode gap, NV$_1$, located 16\,$\upmu$m away along the $y$-axis and a third, NV$_2$, located 50\,$\upmu$m away along $y$, as shown in Fig.~\ref{fig:3}(d). All three NV centres are at approximately the same depth of 4$\,\upmu$m. We then pumped a given NV centre with 3.5\,mW of green light for a variable duration before returning to the same optimal electrode readout position and reading the trapped charge. Repeating for all three defects, we measured the accumulated charge $Q_\text{int}$ as a function of pumping time in Fig.~\ref{fig:3}(d). We observed that the two NVs located further away take longer to register a measurable charge on the electrodes during readout, though even at 50\,$\upmu$m away can still be detected with sufficient pumping time. Interestingly, after such long pump times, the liberated charge at the electrodes is essentially the same as for the two closer NVs, suggesting that trapped charge in the bulk of the diamond is \emph{not} contributing significantly to the measured photocurrent transient. The same fixed number of electrode traps addressed by the laser appear to fill more slowly for more distant charge sources, but release the same amount of charge upon illumination. The significantly larger region of bulk sample exposed to carrier flux from the distant source NV either does not trap significant numbers of carriers, or is not depleted upon initation of the EGPC readout. We return to this point again later in the manuscript.           


\subsection{Photocurrent imaging}

\noindent Having established that the charge storage and readout mechanism relies on metal-diamond interface states with photoactivation energies around 2.2\,eV, we now explore in more detail the identity of the trapped charge carriers. We employ the array of NV$^-$ defects as single-pixel hole-current detectors~\cite{wood_3d-mapping_2024} that change charge states (and hence emission wavelength) upon capture of a hole. With appropriate optical filtering (long-pass cutoff wavelengths of 650\,nm), NV$^-$ appears bright while NV$^0$ appears dark under weak orange (3\,$\upmu$W, 594\,nm) illumination. Single-shot charge state identification of single NVs amounts to determining the photons emitted in a 50\,ms pulse of orange light. Fluorescence histograms were measured for several dozen NV centres in proximity to the electrode assembly. We calibrated each histogram using short, strong red (633\,nm) illumination to create NV$^0$ with almost unity probability~\cite{wood_wavelength_2024} and green light which yields NV$^-$ with $\sim70\%$ probability~\cite{wirtitsch_exploiting_2023}, and characteristic NV$^-$ and NV$^0$ count thresholds were described by bi-modal Poissonian distributions.\newline

\begin{figure*}[t]
    \centering
    \includegraphics[width=\textwidth]{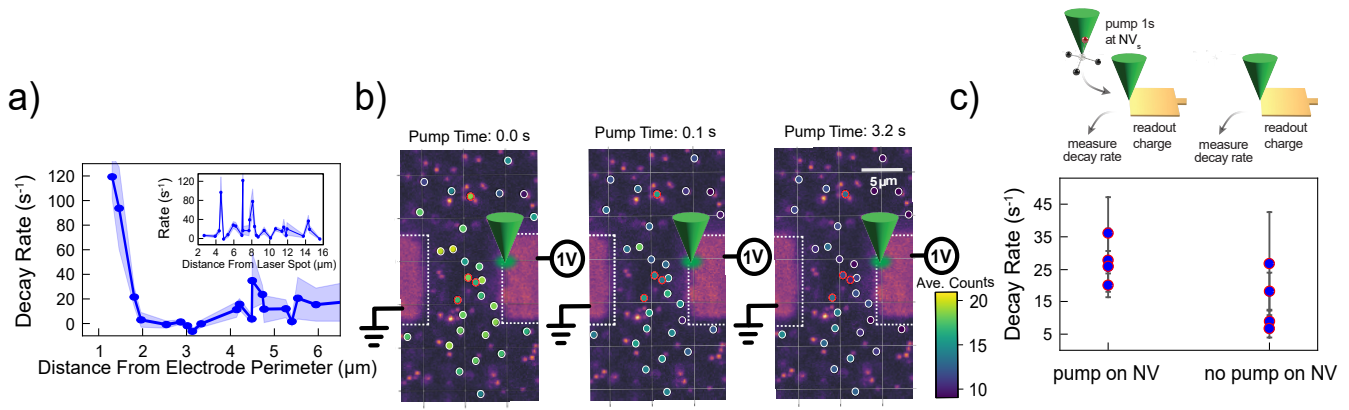}
    \caption{\textbf{Spatial mapping of hole current density.} a) Decay rate from NV$^-$ to NV$^0$ measured as a function of the shortest distance from the electrode edge to each NV centre. Inset: Decay rates measured as a function of radial distance from the contact illumination spot. b) Maps of average photon counts obtained from each measured NV centre (circled areas) as a function of contact illumination time under a 1\,V bias. PL image was acquired with green light, and the electrode boundaries are shown in dashed white lines. Transition from green (NV$^-$) to blue (NV$^0$) indicates increased probability of hole capture. The laser illumination spot is indicated by the green cone. c) Decay rate from NV$^-$ to NV$^0$ for the case where the NV is first pumped for 1~s to generate photocarriers prior to EGPC initiation (left) and the case where no photocarriers are introduced before EGPC initiation (right).}
    \label{fig:4}
\end{figure*}

\noindent The target NV centres were then prepared in the NV$^-$ (bright) state using green light and then the laser (3.5\,mW) was moved to the electrode boundary to stimulate an EGPC under an applied 1\,V bias. After current generation for some time $T$, the charge state of each NV was read out to determine if any defects underwent charge transfer to NV$^0$ due to photogenerated hole capture. We found that a hole current emanated from the entire electrode junction, which resulted in a sharp dependence of the hole capture rate measured as a function of shortest distance from the electrode boundary, shown in Fig.~\ref{fig:4}(a). As the illumination time at the contact increases, the array of NV$^-$ centres become progressively darker due to higher probability of hole capture, as illustrated in Fig.~\ref{fig:4}(b). Notably, the spatial dependence of hole capture rate is lost when considering the laser spot to be a point source of holes. As shown in the inset of Fig.~\ref{fig:4}(a), defects closer to the electrode, and not the point of illumination on the electrode, are more likely to capture holes. The behavior of the NV$^-$ to NV$^0$ decay indicates that the current generated by the EGPC protocol is injected into the diamond non-locally to the laser illumination spot along the entire diamond-metal interface. Next, we pumped the source NV with green light to generate charge carriers that become trapped at the electrode boundary and repeated the above experiment. Here, the additional trapped charge from the NVs -- which can thus be identified conclusively as holes -- result in a faster decay in NV$^-$ probability (shown in Fig~\ref{fig:3}(c) by a representative number of NV centres highlighted in red circles in Fig~\ref{fig:3}(b)), consistent with additional photogenerated holes being liberated from the electrode charge traps. 

\section{Discussion}
\noindent We have introduced a novel photoelectric detection scheme where charge carriers generated by single NV centres can be stored almost indefinitely in long-lived charge traps primarily localized at metal-diamond Schottky junctions. As the NV photoionization process is spin-dependent, the charge flux emanating from a single NV is also spin dependent. We confirmed that our technique facilitates measurement of single NV spins under coherent control protocols, highlighting its potential as an alternative photoelectrical interface for quantum sensing and information processing. Our method has several attractive features, including long-term storage of spin information in charge traps, no crosstalk between microwave control fields and electrical readout, and no requirement for lock-in detection schemes to facilitate electrical readout of pulsed measurements.\newline

\noindent There is significant scope to build upon our findings. A necessary avenue of future work is to identify in detail the traps responsible for carrier trapping, which our work suggests are located at the interface between the metal electrode and diamond, have an activation energy near 2.2\,eV, and are predominantly acceptor-like given their propensity to trap and emit holes. The Cr/Au electrodes used in this work were deposited onto an oxygen-terminated surface with no subsequent annealing. Future work could experiment with alternative contact materials (e.g., Ti, Al, graphite~\cite{forneris_mapping_2018}), surface terminations (O, H), and processing steps such as annealing to create more Ohmic-like electrode contacts \cite{broadway_spatial_2018}. Further, varying the electrode geometry or fabricating circular electrodes proximal to/surrounding single NV centres could also serve to improve trapping yield. Conversely, the traps identified in this work could be responsible for reduced collection efficiency or unwanted background current in conventional photoelectric detection~\cite{hruby_magnetic_2022, rieger_fast_2024}, so combining materials engineering with careful analysis of trapping behaviour using the techniques reported here could also bring benefits.\newline

\noindent Our method does not confer any immediate advantages in spin contrast over existing PD or spin-to-charge conversion schemes, as ultimately the photophysics of the NV and specifically the inter-system branching ratio for $m_S = \pm1$ states limits contrast to $\sim30$-$40$\%~\cite{hanlon_enhancement_2023}. However, it is worth noting that combining spin-to-charge protocols~\cite{shields_efficient_2015, jayakumar_spin_2018, rovny_nanoscale_2022} that optimise spin-selective ionisation with electrical readout has not been explored significantly. The fast pulsed nature of these sequences (using red ionisation pulses 10\,mW$+$ of only a few 10\,s of ns) may be difficult to implement in conventional PD, whereas such protocols are eminently feasible and compatible with our technique~\cite{jayakumar_long-term_2020}.\newline

\noindent Beyond the proof-of-principle stage, we can consider intriguing extensions that realize the full potential of hybrid photoelectrical readout. One particularly interesting possibility is to increase PD bandwidth by releasing the trapped charge faster using ultrafast laser pulses. Fast, sensitive charge-coupled amplifiers with sensitivities into the fC range and bandwidths of GHz or more~\cite{portier_carrier_2023} could then be employed in place of slow, high gain transimpedance amplifiers ubiquitous in PD. Using a second laser focused only on the optimum electrode readout point~\cite{rovny_nanoscale_2022, goldblatt_quantum_2024} may also enable continuous readout of charge accumulation during NV measurements, precluding the need to accumulate sufficient charge over long integration times to yield measurable charge transients. As we found that hole currents originate from the entire metal-diamond interface region rather than the illuminated region alone, it may also be possible to realize fully integrated PD by generating these carriers remotely by e.g., extending metal contacts further away from the NV centres under study and situating small on-chip LEDs proximal to the far edges of those contacts. Optical illumination may also be unnecessary - for instance, our measurements suggest the possibility that generation of a current alone may be sufficient to liberate the NV-sourced charges trapped at the metal interface. A natural extension could then be to trigger a current via electrical breakdown of the device by increasing bias voltages to $\sim100$-$200$\,V, with the breakdown voltage $V_B$ possibly being set by the trapped NV charge. Such a scheme would realise room-temperature, all-electrical readout method with potentially superior sensitivity and bandwidth to conventional PD.\newline

\noindent In conclusion, we have shown that interface traps at a diamond-metal Schottky junction provide a means to probe single-spin dynamics with performance comparable to conventional PD schemes. This work opens numerous avenues of further inquiry targeted towards both the fundamental mechanism of EGPC readout as well as the best pathways towards engineering practical integrated devices for rapid, high-fidelity spin measurements. We anticipate that optimisation of the electrode geometry and metallisation chemistry will yield substantial improvements in performance, while deployment of dedicated spin-to-charge conversion protocols could improve spin-dependent carrier generation.

\section*{Acknowledgments}
\noindent This work was supported by the Australian Research Council (ARC) through grant DE210101093 and the ARC Centre of Excellence in Quantum Biotechnology through grant CE230100021. D.J.M. acknowledges support through a University of Melbourne McKenzie Fellowship.

\end{document}